\documentclass[a4paper,showpacs,amsmath,showkeys]{revtex4}
\usepackage[latin1]{inputenc}
\usepackage{graphicx}
\usepackage{graphics}
\usepackage{epsfig}
\usepackage{amsmath}
\begin{document}
\newcommand{\be}{\begin{equation}}
\newcommand{\ee}{\end{equation}}
\title{Bright solitons in asymmetrically trapped Bose-Einstein condensates}
\author{Sk.Golam Ali, B.Talukdar}
\email{binoy123@sancharnet.in}
\author{S.K.Roy}
\affiliation{Department of Physics, Visva-Bharati University,
Santiniketan 731235, India}
\begin{abstract}
  We study the dynamics of bright solitons in a Bose-Einstein condensate (BEC) confined in a highly asymmetric trap.  While working within the framework of a variational approach we carry out the stability analysis of BEC solitons against collapse. When the number of atoms in the soliton exceeds a critical number $N_c$, it undergoes the so called primary collapse. We find an analytical expression for $N_c$ in terms of appropriate experimental quantities that are used to produce and confine the condensate. We further demonstrate that, in the geometry of the problem considered, the width of the soliton varies inversely as the number of constituent atoms. 
\end{abstract}

\maketitle
\section*{\large{1. Introduction}}
A Bose-Einstein condensate (BEC) consists of trapped ultracold atoms all in the same quantum state. In this state the atoms lose 
their individual identities and behave as a single collective wave which is large enough to be optically imaged. In order to creat a BEC, atoms are first
confined within a strong magnetic field and then the temperature of the atomic gas is continually lowered by laser and 
evaporative cooling until the condensate is formed. If one confines the BEC in only two directions, it will tend to disperse in 
the free direction. Because of the energetics involved, the atom-atom interaction in freely propagating BEC is 
characterized by the s-wave  scattering length.
The Feshbach resonance \cite{1} allows one to continuously tune the scattering length from a positive to negative value 
(repulsive to attractive  interaction ) by means of applied magnetic field. For attractive atomic interaction we can have 
coherently propagating matter-wave packets which travel over the BEC with neither attenuation nor change in shape. 
These are the so called bright solitons.  For repulsive interaction we shall have dark solitons. 
Understandably, a bright soliton is a peak on the BEC while a dark soliton is a notch with a characteristic phase step across it. 
In a BEC of $^7Li$ atom Rice and Paris teams \cite{2} produced
bright solitons, each of which represents a condensate of actual atoms extracted from the main BEC. 
\par 
When the number of atoms in a bright soliton exceeds a critical value, it becomes unstable due to focusing nonlinearity
arising from the attractive atom-atom intarection. The transverse dimensions of the confinement then cause the soliton to collapse. This is often referred to as the primary collapse \cite{3}. In the present work we shall envisage a variational study for the stability of bright soliton in a highly elongated trap and thereby calculate the critical number of atoms ($N_c$) 
that a soliton can hold before it undergoes the so called primary collapse. We shall see that the merit of our approach is its directness and simplicity because the variational method sought by us provides a straightforward analytical model to understand the dynamics
of bright solitons. 
\par 
To extract the relevant physical information regarding stability and / or collapse we shall work within the framework of a mean field approximation. In this approximation the dynamics of a BEC is modelled by 3D Gross-Pitaevskii (GP) equation 
\be
i\hbar\frac{\partial \psi}{\partial t }=-\frac{\hbar^2}{2m}\nabla^2\psi+V(\vec{r})\psi+U_0|\psi|^2\psi,
\ee
where $\psi(\vec{r},t)$ is the macroscopic wave function of the condensate. This wave function is also  called the 
order parameter. Here $U_0=\frac{4\pi \hbar^2 a_s}{m}$ represents the interatomic interaction with $ a_s$, the two-particle $s$-wave 
scattering length and $m$, the mass of the atom. The wave function $\psi$ is normalized to the number of particles $N$ in the 
condensates such that
\be
\int|\psi|^2 d\vec{r}=N.
\ee
The potential $V(\vec{r})$ confines the atoms in a trap. For harmonic trapping, $V(\vec{r})$ is given 
 by
\be
V(\vec{r})={1\over 2} m \nu^2\left(\lambda^2_x x^2 +\lambda^2_y y^2+\lambda^2_z z^2\right).
\ee
The parameters $\lambda_x$, $\lambda_y$ and $\lambda_z$ describe anisotropy of the trap in the $x$, $y$ and $z$ direction respectively.
We shall work with a highly asymmetric trap as determined by $\lambda_x=\lambda_y=1$ and $\lambda_z=\frac{\nu_z}{\nu}\ll 1$.
Here $\nu_z$ represents the frequency along the $z$ directions and $\nu=\nu_r$, the radial frequency. Our system of interest is thus
a quasi-one dimensional (Q1D) BEC dispersing along the z direction. In the following we derive an appropriate version of the GP equation that will be useful to study the dynamics of BEC in highly asymmetric traps.
\par 
We consider $(1)$ in a geometry in which the trapping potential in $z$ is much weaker than the corresponding 
potential in $r=(x^2+y^2)^{1\over2}$. Further, we write the equation in terms of dimensionless variables defined by 
\begin{eqnarray}
\tau=\nu t,&\rho={r\over a_0},&
s={z\over a_0},\,\,\,\,
\psi(r, z,t)= {u(\rho,s,\tau)/ a_0^{3\over 2}}.
\end{eqnarray}
This gives 
\be 
iu_{\tau}+{1\over 2}\nabla^2 u-{1\over 2}\left(\rho^2+\lambda_z^2 s^2\right)u-\frac{4\pi a_s}{a_0}|u|^2u=0.
\ee
Here $a_0=\sqrt{\frac{\hbar}{m \nu}}$ is the size of the ground state solution of the noninteracting  GP equation. It is obvious that
\be 
\int|u|^2 d^3\rho = N.
\ee
We assume a separable ansatz for the solution of $(5)$  such that \cite{4}
\be
u(\rho,s,\tau)=\phi(\rho)\xi(s,\tau).
\ee
From $(5)$ and $(7)$ we have
\begin{eqnarray}
 {1\over {\xi}} \left(i\xi_{\tau}+{1\over 2}\xi_{2s}-{1\over 2}\lambda_z^2 s^2\xi\right)-\frac{4\pi a_s}{a_0}|\xi|^2|\phi|^2 
 \nonumber\\={1\over {\phi}}\left(-{1\over 2}\nabla^2_{\rho}\phi+{1\over 2}\rho^2\phi\right),
 \end{eqnarray}
where  $\nabla^2_{\rho}$ stands for the Laplacian in the radial coordinate. In $(8)$ the subscripts on $\xi$ stands for
partial derivative with respect to that particular independent variable. More specifically,
$\xi_{2s}=\frac{\partial^2 \xi}{\partial s^2}$.  This equation shows that the presence of atom-atom interaction 
does not permit clearcut separation of variables. However, the fourth term in equation $(8)$ is quite small. 
Thus, $ \phi$ may be assumed to satisfy
\be
-{1\over 2}\nabla^2_{\rho}\phi+{1\over 2}\rho^2\phi=\nu_{\rho}\phi
\ee
with $\nu_{\rho} $ being related to $\nu_{r}$ by a scale factor determined by the change of variables sought in $(4)$.
Equation $(9)$ represents the well-known eigenvalue problem for the two dimensional harmonic oscillator with the ground sate 
solution given by
\be
{\phi_0}(\rho)={\rm e}^{-\rho^2/2}.
\ee
Combining $(8)$ and $(9)$ we write 
\be
 i\xi_{\tau}+{1\over 2}\xi_{2s}-{1\over 2}\lambda_z^2 s^2\xi-\frac{4\pi a_s}{a_0}|\xi|^2|\phi|^2\xi
 ={\nu_{\rho}}\xi.
\ee
The low-frequency vibration along the $z$ direction is quite unlikely to excite the two dimensional bosonic oscillator from its ground 
state. Thus $(11)$ can be multiplied by $\phi \phi^{\star}$ and integrated over the $\rho$ coordinate to get

\be
 i\xi_{\tau}+{1\over 2}\xi_{2s}-{1\over 2}\lambda_z^2 s^2\xi-\frac{2\pi a_s}{a_0}|\xi|^2\xi
 ={\nu_{\rho}}\xi.
\ee
Equation $(12)$  represents the GP Equation for a Q1D trap. For a true 1D system one does not expect collapse of the system with increasig number of atoms. But the use of Q1D trap in controlling the condensate motion may result in the collase of a BEC soliton when the number of atoms in it exceeds a critical value, say, $N_c$. We  shall work with $(12)$ to provide an analytical model to study the collapse dynamics of bright solitons in a Q1D trap. Interestingly, $(12)$ can  be written in a more convenient form by using the change of variable 
\be
\xi(s,\tau)=\chi(s,{\tau})e^{-i{\nu_{\rho}}\tau}.
\ee
From $(12)$ and $(13)$ we get
\be
 i\chi_{\tau}+{1\over 2}\chi_{2s}-{1\over 2}\lambda_z^2 s^2\chi-\frac{2\pi a_s}{a_0}|\chi|^2\chi
 =0
 \ee
with
\be
 {\int}^{+\infty}_{-\infty}|\chi|^2 ds=N/{\pi}.
\ee
Equation $(14)$ represents the desired form of the evolution equation in which the atom-atom interaction is characterized
by a negative scattering length. The realistic 1D limit in $(14)$ is not a true  1D system because this equation involves the effect of transverse degrees of freedom through $\lambda_z$ and $a_0$. In this context we note that a similar equation with $\psi(r,z,t)$ chosen as
$u(\rho,s,\tau)/\sqrt{a^3_0/N}$ was used by P\'erez-Garc\'ia et al \cite{4} to qualitatively demonstrate that if the number of particles is large enough, the condensate is unstable and collapse occurs. We are ,however, interested to derive a straightforward analytical model to understand the collapse dynamics and thereby provide a quantitative estimate for $N_c$. To that end we convert, in section 2, the initial-boundary value problem in $(14)$ to a variational problem. In particular, we present an expression for the Lagrangian density and a trial wave function involving variational parameters to study the stability of bright solitons against collapse.  We also obtain the evolution equations for these parameters. In section 3 we judiciously use the derived evolution laws to study the soliton dynamics with particular emphasis on the stability of solitons against collapse.
\section*{\large{2. Variational formulation}}
The action principle
\be
\delta\int\int{\cal L}(\chi, \chi^\star, \chi_s, \chi_s^\star, \chi_{\tau},\chi_{\tau}^\star) ds \,d{\tau}
\ee
with the Lagrangian density given by
\be
{\cal L}={i\over 2} \left(\chi \chi_{\tau}^\star-\chi^\star \chi_{\tau} \right)+{1\over 2}\lambda_z^2 s^2\chi\chi^\star+\frac{\pi a_s}{a_0}
\chi^2{\chi^\star}^2+{1\over 2}\chi_s^\star\chi_s,
\ee
is equivalent to $(14)$. We shall use this expression for ${\cal L}$ to study the dynamics of bright solitons in terms of a variational
method often called the Ritz optimization procedure \cite{5}. In this procedure the first variation of the variational functional is made to vanish
within  a set of suitable chosen trial function such that the field theoretical problem under consideration reduces to a simple problem
of point mechanics. For the negative scattering length the Gaussian trial function for ${\chi(s,\tau)}$ is a very reasonable ansatz. Thus
we write
\be
\chi(s,\tau)= A(\tau) \exp\left[ -s^2/{2 a^2(\tau)}+i b(\tau) s^2/2\right].
\ee 
Here $A(\tau)$ is a complex amplitude, $a(\tau)$ - the width of the distribution and $b(\tau)$ - the frequency chirp. The phase of 
the condensate $\delta(\tau)$ is defined by $A(\tau)=|A(\tau)| {\rm e}^{\delta(\tau)}$. The amplitude $A(\tau)$, width $a(\tau)$  
and the chirp
$b(\tau)$ will all vary with the time parameter $\tau$. The initial condensate at rest will have $da(\tau)/d\tau=0$. 
\par 
Inserting the trial function in $(18)$ into the variational principle stated in $(16)$ we obtain a reduced variational problem
\be
\delta \int\left<{\cal L}\right> d{\tau}=0
\ee
with
\be
\left<{\cal L}\right> = \int^{+\infty}_{-\infty}{\cal L}_G \, ds.
\ee
Here ${\cal L}_G$ represents the result obtained by inserting the Gaussian ansatz $(18)$ into the Lagrangian density in $(17)$. 
It is rather straight forward to perform the integration in $(20)$ and get 
\begin{eqnarray}
{\left<{\cal L}\right>} =\sqrt \pi {i\over 2} \left(A A^\star_{\tau}-A^\star A_{\tau}\right) a+
{1\over 4}b_{\tau}a^3 A A^\star+\nonumber\\{1\over 4}\lambda_z^2 a^3 A A^\star+{\pi \over {\sqrt 2}}{a_s\over a_o}a A^2 {A^\star}^2  +
{1\over 4}\left({1\over a}+b^2a^3\right) A A^\star{\bigg\rgroup}.
\end{eqnarray}
\par Let us now obtain the variational equations for the Gaussian parameters $A(\tau)$, $A^\star(\tau)$, $a(\tau)$ and $b(\tau)$ which 
follow from the vanishing conditions of $\frac{\delta\left<{\cal L}\right>}{\delta A}$,  
$\frac{\delta\left<{\cal L}\right>}{\delta A^\star}$, $\frac{\delta\left<{\cal L}\right>}{\delta a}$ and 
$\frac{\delta\left<{\cal L}\right>}{\delta b}$ . These  equations are given by
\begin{eqnarray}
\frac{\delta\left<{\cal L}\right>}{\delta A}=iA^\star_{\tau}a+{i\over 2}A^\star a_{\tau}+{1\over 4}b_{\tau}a^3 A^\star+
{1\over 4}\lambda_z^2 a^3 A^\star\nonumber\\+{\pi{\sqrt 2}}{a_s\over a_o}a A {A^\star}^2 +{1\over 4}\left({1\over a}+b^2a^3\right) 
A^{\star}=0, 
\end{eqnarray}

\begin{eqnarray}
\frac{\delta\left<{\cal L}\right>}{\delta A^\star}=-iA_{\tau}a-{i\over 2}A a_{\tau}+{1\over 4}b_{\tau}a^3 A+
{1\over 4}\lambda_z^2 a^3 A\nonumber\\+{\pi{\sqrt 2}}{a_s\over a_o}a A^2 {A^\star} +{1\over 4}\left({1\over a}+b^2a^3\right) A=0, 
\end{eqnarray}

\begin{eqnarray}
\frac{\delta\left<{\cal L}\right>}{\delta a} ={i\over 2} \left(A A^\star_{\tau}-A^\star A_{\tau}\right) +
{3\over 4}b_{\tau}a^2 A A^\star+\nonumber\\{3\over 4}\lambda_z^2 a^2 A A^\star +{\pi \over {\sqrt 2}}{a_s\over a_o} A^2 {A^\star}^2  
+\nonumber\\
{1\over 4}\left(-{1\over a^2}+3b^2a^2\right) A A^\star = 0  
\end{eqnarray} and 

\begin{eqnarray}
\frac{\delta\left<{\cal L}\right>}{\delta b} ={1\over 2}b a^3 A A^\star -{1\over 4}\frac{\partial }{\partial \tau}(a^3 A A^\star)=0.
\end{eqnarray}
From $(22)$ and $(23)$ we have found 
\be
\frac{d }{d \tau}\left(a A A^\star\right)=0
\ee
such that
\be
a|A|^2={ Q},\,\,{\rm {a\,\, constant}}.
\ee
The constant ${Q}$ is simply related to the number of particles in the condensate since the value of the integral $(15)$
is ${{\sqrt\pi} a|A|^2}$. Combining $(25)$ and $(27)$ we get 
\be
b=\frac{d }{d \tau}(\ln\,a).
\ee
Equations $(27)$ and $(28)$ clearly show that if we can derive a method to calculate the values of $a(\tau)$, the other parameters of the 
condensate will be automatically determined. Fortunately, $(22)$, $(23)$, $(24)$  and $(28)$ can be combined to write a second-order 
ordinary
differential equation, the first integral of which gives
\be
{1\over 2}\left(\frac{da}{d\tau}\right)^2+{1\over 2}\lambda_z^2 a^2+{\sqrt{2\over \pi}}\frac{N a_s}{a_0}{1\over a}+{1\over 2a^2}=E,
\ee
with $E$, the constant of integration.
\par The equation for $a(\tau)$ in $(29)$ is related to the motion of a particle in a potential field $V(a)$ so that  
\be
{1\over 2}\left(\frac{da}{d\tau}\right)^2+V(a)=E.
\ee
Here
\be
V(a)={1\over 2}\lambda_z^2 a^2+{P\over a}+{1\over 2a^2}\, ,\,\,P={\sqrt{2\over \pi}}\frac{N a_s}{a_0}.
\ee
Thus one would like to interpret the constant of the motion $E$ as the total energy of the particle. It is easy to verify that there is no physical uncertainty in the identification sought because $\int_{-\infty}^{+\infty}{\cal H}_G ds$ represents the left side of $(29)$.Here  ${\cal H}_G$ stands for the result of the Hamiltonian density calculated from $(18)$ and rewritten by using the Gaussian ansatz. Obviously, $E$ is determined by the 
initial conditions of the second order differential equation from which $(30)$ has been extracted. It is not difficult to solve $(30)$
and look for the dynamics of the condensate. However, the analysis of the equilibrium point obtained from the extrimum of $V(a)$ 
written as 
\be
\frac{dV(a)}{da}=0
\ee
can give some illuminating results.
\section*{\large{3. Dynamics of bright solitons }}
For bright solitons the nonlinear interaction is attractive and the scattering length $a_s<0$. In this case we shall use $P=-|P|$
and carry out the subsequent analysis by using only the numerical values of $a_s$. We shall make use of $(32)$ to derive a simple physical picture for the collapse dynamics of bright solitons when the trap of the BEC is relaxed in one direction.
From $(31)$ and $(32)$ with $P=-|P|$  we get 
\be
\lambda_z^2 a^4+|P|a-1=0.
\ee
The equilibrium point determined by $(33)$ should be a minimum for $(14)$ to support a soliton solution. This gives
\be
\lambda_z^2 a^4-2|P|a+3=\beta a^4,\,\,\,\beta > 0.
\ee
Eliminating $|P|$ from $(33)$ and $(34)$ we find that
\be
a=\left(\frac{1}{\beta-3\lambda_z^2}\right)^{1\over4}, \,\,\,\beta>3\lambda_z^2
\ee
is a particular solution of $(33)$ and $(34)$. From $(35)$ and $(33)$ or $(34)$ we get 
\be
|P|=\frac{\left(\beta-4\lambda_z^2\right)}{\left(\beta-3\lambda_z^2\right)^{3\over4}}.
\ee
The form of $(36)$ imposes a further restriction on the values of $\beta$ than that given in $(35)$ and sets a lower bound for it.
Using $\beta=\gamma \lambda_z^2$ we write $(36)$ in the form 
\be
|P|=\frac{\left(\gamma-4\right)}{\left(\gamma-3\right)^{3\over4}}\sqrt{\lambda_z}.
\ee
Thus non zero values of $P$ will be obtained for $\gamma>4$ only.  For $\gamma=4$ the interaction term vanishes and GP equation becomes
linear and soliton formation becomes impossible.
\par
From $(31)$ and $(37)$ we obtain an expression
\be
N=\sqrt{\pi\over2}\frac{a_0}{|a_s|}\frac{\left(\gamma-4\right)}{\left(\gamma-3\right)^{3\over4}}\sqrt{\lambda_z}
\ee 
 for the number of atoms in the quasi-1D soliton.  

\begin{figure}
\includegraphics[width=0.40\columnwidth, angle=-90] {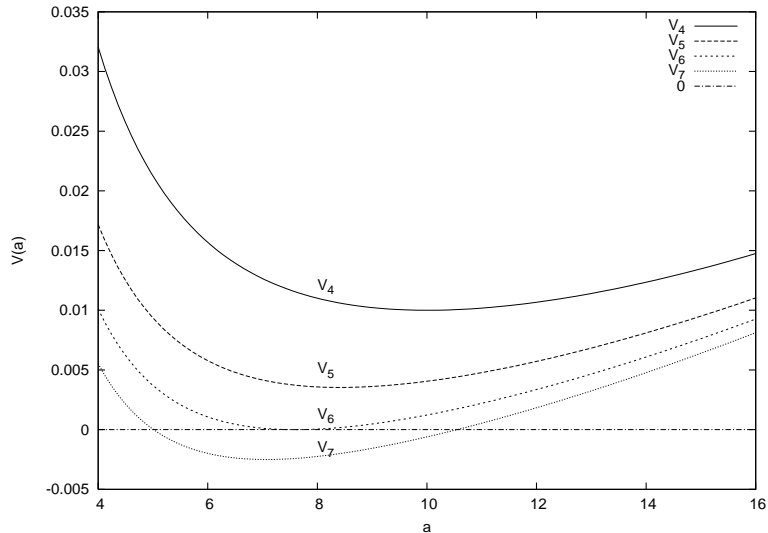}
\caption{The potential $V(a)$ as a function of $a$ for $\lambda_z={1\over100}$}
\end{figure}
\par In Fig. 1 we plot the potential $V(a)$ in $(31)$ as a function of $a$ for $a_s=-1.59\times10^{-4}\mu m$,
the scattering length of $^7Li$ as used in the experiment of Strecker et al. \cite{6}. We have chosen to work with $\lambda_z=\frac{4}{400}$. In this figure we have four curves represented by $V_4(a)$, $V_5(a)$, $V_6(a)$
and $V_7(a)$ corresponding to $\gamma=4,\,\,,5,\,\,6,\,\, {\rm{and}}\,\, 7$ repectively. A common feature of all these potentials is that each of them exhibits a minimum. The  curve for $V_7(a)$ represents a potential well between $a_1=5.0114$ and $a_2=10.5468$. The minimum of the well is negative. A mechanical analogy suggests  a solution which oscillates between the zeros of $V_7(a)$. In this case, spreading of the BEC is stopped at $a=a_2$ by nonlinear effects which subsequently compress the BEC back to the initial width. This behavior is repeated in an oscillatory manner. In this situation the BEC soliton will become unstable and lead to a mechanical collapse \cite{7}. A similar situation arises for other  values of $\gamma > 6$.

\par
 For $\gamma=6$ the potential well degerates into a single point such that $V_6(a)$ touches the a axis at a particular point where the potential has a stable minimum. Understandably, a particle releasesd at this point
 will stay there. In the present context this implies that for our chosen value of $\lambda_z$ and $\gamma=6$   the BEC bright soliton will be critically stable. Using $\gamma=6$ we get 
\be
 N_{c}=0.8774\sqrt\frac{\pi}{2}\frac{a_0}{|a_s|}\sqrt{\lambda_z}.
 \ee  
After the number of atoms exceeds this critical number the soliton  becomes unstable.
More than a decade ago Ruprecht et al. \cite{8} used a purely numerical routine to set a limit for the critical number of atoms after which the BEC with attractive two-body interaction becomes unstable. This limit was examined by Gammal et al. \cite{9} for different trapping geometries. Interestingly, the analytical expression
 in $(39)$ is in agreement with the observations of refs. $8$ and $9$. From $(39)$ it is clear that $\frac{N_c |a_s|}{a_0}\ll1$. This represents the well-known relation for the existance of stable solitons \cite{10}. 
\par
It will be interesting  to see what happens if $\gamma < 6$. In order that we look at the curve represented by $V_4(a)$. For $\gamma=4$, $|P|=0$ and we do not have a nonlinear term in the GP equation. In this case, no soliton can be formed. The minimum of $V_4$ is positive. This appears to  suggest that weaker nonlinearies leading to potential curves with positive minima will not be able to produce matter-wave bright solitons. In our figure $V_5(a)$ represents one such curve.
\par 
Equations $(35)$ and $(37)$ can be combined to write

\begin{equation}
a(\tau)=\frac{C}{N},
\end{equation} 
where
\begin{equation}
C=\sqrt{\frac{\pi}{2}}\left( \frac{\gamma-4}{\gamma-3}\right) \frac{a_0}{|a_s|}.
\end{equation} 
The result in $(40)$ is remarkable and implies that, at a given instant of time, if the number of atoms in the soliton
increases, it becomes narrower. It is a real curiosity to note that the result in $(40)$ holds good even in
the absence of trapping \cite{4}

\section*{\large{Acknowledgements}}
 One of the authors (S.G.A) is thankful to the UGC, Govt. of India for a Research Fellowship.

 \end{document}